\documentclass[sigconf]{acmart}

\usepackage{tikz}

\usepackage{multirow}
\usepackage{array}
\usepackage{algorithm, algpseudocode}
\usepackage{algorithmicx}
\usepackage{tabularx}
\usepackage{booktabs}

\usepackage[frozencache,cachedir=.]{minted}

\usepackage{url}

\usepackage{threeparttable} 
\usepackage{textcomp}
\usepackage{hhline}
\usepackage{stfloats}
\usepackage{float}

\newcommand{\emu}[0]{{\sc Ember-IO}}
\newcommand{\ptwoim}[0]{P$^{2}$IM}
\newcommand{\fcov}[0]{FERMCov}

\usepackage{etoolbox}
\usepackage{comment}

\newcommand{\circled}[2][]{
  \tikz[baseline=(char.base)]{%
    \node[anchor=text, shape=circle,draw, inner sep=0pt, minimum size=0.5em] (char){#1\strut};
    \node at (char.center) {\makebox[0pt][c]{#2}};}}
\robustify{\circled}

\AtBeginDocument{%
  }

\setcopyright{acmcopyright}
\copyrightyear{2023}
\acmYear{2023}
\acmDOI{XXXXXXX.XXXXXXX}

\acmConference[ASIA CCS'23]{18th ACM ASIA Conference on Computer and Communications Security}{July 10--14,
  2023}{Melbourne, Australia}
\acmPrice{15.00}
\acmISBN{978-1-4503-XXXX-X/18/06}

\begin{document}

\title{{\textit{Ember-IO}}: Effective Firmware Fuzzing with Model-Free Memory Mapped IO}

\author{Guy Farrelly}
\email{guy.farrelly@adelaide.edu.au}
\affiliation{\institution{The University of Adelaide}
\country{Australia}}
\author{Michael Chesser}
\email{michael.chesser@adelaide.edu.au}
\affiliation{\institution{The University of Adelaide}
\country{Australia}}
\author{Damith C. Ranasinghe}
\email{damith.ranasinghe@adelaide.edu.au}
\affiliation{\institution{The University of Adelaide}
\country{Australia}}

\begin{abstract}

Exponential growth in embedded systems is driving the research imperative to develop fuzzers to automate firmware testing to uncover software bugs and security vulnerabilities. But, employing fuzzing techniques in this context present a uniquely challenging proposition; a key problem is the need to deal with the \textit{diverse} and \textit{large number} of peripheral communications in an automated testing framework. Recent fuzzing approaches: i)~employ re-hosting methods by executing code in an emulator because fuzzing on resource limited embedded systems is slow and unscalable; and ii)~\textit{integrate models of hardware behaviour} to overcome the challenges faced by the massive \textit{input-space} to be explored created by peripheral devices and to generate inputs that are effective in aiding a fuzzer to make progress.

Our efforts expounds upon program execution behaviours unique to firmware to address the resulting input-space search problem. The techniques we propose improve the fuzzer’s ability to generate values likely to progress execution and avoids time consumed on mutating inputs that are functionally equivalent to other test cases. 

We demonstrate the methods are highly efficient and effective at overcoming the input-space search problem. Our emulation-based implementation, \emu{}, when compared to the existing state-of-the-art fuzzing framework across 21 firmware binaries, demonstrates up to 255\% improvement in blocks covered. Further \emu{} discovered 6 new bugs in the real-world firmware, previously not identified by state-of-the-art fuzzing frameworks. Importantly, \emu{} integrated with the state-of-the-art fuzzer, {\sc Fuzzware}, demonstrates similar or improved coverage across all firmware binaries whilst reproducing 3 of the 6 new bugs discovered by \emu{}.
\end{abstract}


\maketitle
\section{Introduction}

Advances in micro-controller units (MCUs) has resulted in the proliferation of embedded devices in a drive to create smarter and connected things. These devices are resource limited and have only a fraction of the security features of typical desktop machines. Common errors that cause a program to terminate on a desktop application such as null pointer dereferences, stack or heap buffer overflows and double frees do not terminate on many embedded devices, instead execution continues with corrupted memory~\cite{WYCINWYC}. This is unsurprising given the software is often a large single block executing without the supervisory support from an operating system---\textit{a monolithic firmware}. Additionally, these devices are known to have limited security testing and patch management~\cite{iot_security}. We are motivated to consider the problem of developing effective and automated methods for testing and identifying security vulnerabilities of such monolithic firmware.

Fuzz testing, or fuzzing is a common method for automated testing for security bugs. Approaches for firmware execution typically require either the embedded device to be connected to the fuzzer (hardware-in-the-loop), or an emulator. Using physical hardware scales poorly, as each simultaneous fuzzing instance requires an additional device and performance is limited by the typically slow processors in embedded devices. Re-hosting methods or emulation-based approaches are decoupled from the physical hardware as firmware is executed in an emulator such as QEMU~\cite{qemu}. This approach improves scalability and allows for execution on high performance processors. However, emulation-based fuzzers are generally designed for testing desktop applications and cannot be directly used for fuzzing embedded systems, as input is not taken through traditional methods such as files or \texttt{stdin}, but rather as input from peripheral devices such as sensors, actuators or timers. 

Existing methods for embedded systems~\cite{firmadyne,firmafl,firmfuzz,firmae,equafl} provide solutions targeted at the operating system (OS) level for devices using Linux-based systems, but these cannot be applied to monolithic firmware on microcontrollers; often operating without the supervisory control of an operating system. Further, emulation-based approaches for monolithic firmware depend on having an accurate execution environment, including providing values for emulated peripherals. Hence, some fuzzing frameworks \textit{emulate peripherals} using manual software models at a higher abstraction layer~\cite{HALucinator,Li_2021}---while recent methods have investigated approaches to \textit{remove the need for manually written software models}~\cite{uEmu, p2im, fuzzware}.

\subsection{Problem}\label{sec:problem}
Attempting to re-host and fuzz monolithic binaries creates \textit{a search space problem} because of the need to provide fuzzing inputs to a large number of peripheral registers to reach new code or program state. The millions of possible values for each register read and potentially thousands of register reads through a program's execution renders it difficult to effectively mutate inputs to reach new code. Hence, recent approaches for automating re-hosting, \ptwoim~\cite{p2im} and {\textmu}Emu~\cite{uEmu}, attempt to reduce this search space by providing categorisations of registers. Data registers are sourced from the fuzzing input, while the other categories of registers are provided with fixed values or treated as memory--these \textit{non-data categorised register accesses are effectively excluded from the input search space}. This presents a new problem when miscategorisations occur as techniques developed rely on approximate heuristics to generate such models. Further, by removing the register from the fuzzable space, exploration can be limited if valid values are not presented. One example consequence occurs when fixed status values prevent the execution of error handlers. In contrast to explicitly modelling peripheral register types, the most recent method, {\sc Fuzzware}~\cite{fuzzware}, directly reduces the search space. {\sc Fuzzware} utilizes \textit{symbolic execution} to determine the  bits within a register that can influence firmware control flow to allow it to precisely model the necessary bits in the register to fuzz.

\subsection{An Overview of Our Approach}

We propose an effective and simpler alternative to the search space problem. We focus on improving the fuzzer's ability to generate values likely to progress execution (overcome hurdles in firmware initialisation and reaching deep code paths) and avoid time consumed on mutating inputs that are \textit{functionally equivalent} to other fuzz test cases. We briefly describe observations leading to the key techniques we propose, below.

\begin{enumerate}
    \item We observed very high path counts on firmware with existing fuzzing frameworks; this is despite the relatively small program size or block counts. The disparity is due to existing coverage instrumentation methods failing to represent code-coverage when applied to monolithic firmware. Thus, a fuzzer's input generation method informed by code coverage feedback expends \textit{unnecessary effort searching the large input-space without making progress} to explore hitherto unseen code. Notably, existing works~\cite{binaryonly_instrumentation,collafl, fullspeedfuzz, tortoisefuzz, aflsensitive} in the desktop application space have provided evidence of the importance of providing high-quality instrumentation for coverage guided fuzzers. Therefore, we propose a new code coverage instrumentation technique to allow for more accurate identification of truly unique paths we refer to as \textit{Functionally Equivalent Coverage InstRuMentation
(\fcov{})}. Our approach is motivated by the insight that control flow through an embedded program is often interrupted by hardware triggered events, behaviour not typical in desktop applications. Therefore, we propose separating the coverage instrumentation for interrupt control flow from the rest of the firmware control flow.
    \item Recent studies~\cite{p2im,uEmu} have identified the need for certain registers to repeatedly report specific values in order to execute firmware correctly and chosen to fix register values (use \textit{immutable} inputs). But, fixing the values of registers effectively \textit{over restricts the search-space} and can prevent testing of error handlers; importantly, error handlers were identified as a common source of bugs~\cite{ifizz,fifuzz}. Expounding upon this insight we propose the \textit{peripheral input playback} technique to: i)~exploit the knowledge that many peripheral registers often repeatedly return the same value under typical execution conditions; and ii)~benefit from mutated inputs to overcome restrictions on triggering of error handlers.
\end{enumerate}

\subsection{Our Contributions}
Through our efforts, we make the following contributions in this study:

\begin{itemize}
 
\item
    We introduce \textit{functionally equivalent coverage instrumentation} (\fcov{}) and \textit{peripheral input playback} (PIP) techniques to improve the performance of monolithic firmware fuzzers by facilitating a more effective search through the input-space to make progress---overcoming hurdles imposed by MMIO accesses and reaching deep code paths. The techniques we propose provide a \textit{different approach} to current model based methods and can be understood as addressing the problems arising from the large input search space created by the diverse and large number of memory mapped peripherals in firmware fuzzing.
\item
     We implement our techniques in \emu---a highly effective fuzzing framework realised in AFL++-QEMU that \textit{simplifies} re-hosting based fuzzing of monolithic firmware whilst providing a \textit{new}, model-free approach for fuzzing.
\item
    We demonstrate the techniques we propose are generalizable and highly effective in experiments with a diverse set of 21 real-world binaries as a benchmark, and comparisons with the existing state-of-the-art re-hosting method for monolithic firmware---{\sc Fuzzware}. When compared to {\sc Fuzzware}, \emu{} provided up to 255\% improvement in blocks covered, and discovered 6 bugs in the real-world firmware not previously uncovered by existing embedded device firmware fuzzers.
\end{itemize}
We \textbf{open source}\footnote{GitHub Repo: \url{https://github.com/Ember-IO/Ember-IO-Fuzzing}} \emu{}, the experimental datasets, and bug analysis to support reproducibility, further improvements and advance the field of firmware fuzzing.

\section{Background}
Prior to delving into the details of our work, we present a brief background with a focus on highlighting the unique nature of fuzzing embedded systems firmware and briefly revisit coverage guided fuzzing.
\subsection{Embedded Systems}
Embedded systems running on microcontrollers often do not rely on a typical operating system, instead executing a single monolithic program that manages the hardware initialisation, peripheral access, the desired functionality and, in some cases, a minimalistic system to manage tasks. 

\vspace{2mm}
\noindent\textbf{Inputs to Embedded Systems.}~Unlike typical fuzzing targets, inputs to embedded firmware are primarily passed into the program using Memory Mapped Input Output (MMIO) or Direct Memory Access (DMA). MMIO operates by providing a region of memory that directly access the registers of each peripheral device. The registers in this region will typically be one of the following types:

\begin{description}
\item[Control] Used to configure the peripheral device. An example of this is the baud rate of a serial port. 

\item[Status] Used to identify the current state of the peripheral. Bits within a status register may represent whether the peripheral has received new data, or certain errors occurred in the peripheral's operation.

\item[Data] Source or sink of data received or sent by the peripheral. Registers such as these may contain the last reading by an ADC (analog to digital converter), or the byte to be written to a serial port.
\end{description}

While it is common for a register to be used as a single type, in some cases, different bits within a single register may serve different purposes, representing different types. Accesses to memory mapped IO is easily identifiable, as architectures can define designated regions specifically for this purpose. For example, ARM Cortex-M devices reserve the region from \texttt{0x40000000} to \texttt{0x5FFFFFFF} \cite{cortexm4_manual} for peripherals. Registers of each type are scattered throughout this region, with adjacent registers being associated by peripheral rather than type. 

Peripherals may also interact with the MCU using DMA. When interacting using DMA, the peripheral uses a DMA controller to write directly to the memory of the MCU. The data will be present in memory prior to the firmware reading it. This is more common for higher bandwidth peripherals.

An \textit{additional form of input to embedded systems is hardware interrupts}. While they do not provide data, they signal an event has occurred. This causes the processor to jump to an interrupt handler, where the signal is then processed, and the processor returns to the code being executed prior to the interrupt.

In order to fuzz firmware effectively, suitable inputs must be provided from all peripherals used in the loaded firmware to pass initialisation checks. Furthermore, appropriate status register values must be provided to allow continued inputs from data registers. These factors present a uniquely challenging setting for rapid test automation of embedded system firmware.

\subsection{Coverage Guided Fuzzing}
Fuzzers typically take a sample input and progressively mutate this input to trigger different functionality within the program under test. Many fuzzers use knowledge of the code covered in each test to guide the fuzzer to new sections of code. Interesting cases that trigger a new code path are saved to a queue to undergo further mutation. One way in which the code coverage can be tracked is through running the binary inside of an emulator. As the emulator processes each new basic block of instructions, it records this in a map. After executing an input, the fuzzer can analyze this map to determine if there are new entries in the map indicating that new code has been reached.

Popular fuzzers such as AFL~\cite{afl} keep track of the edges visited during program execution to build this map. An edge is a pair consisting of an origin and destination program counter, representing a jump from one block to another. Any test case containing a previously unseen edge will be considered interesting and added as a new path. Test cases that significantly increase the number of times an edge is hit in an execution are also saved for further fuzzing.

\section{Our Approach}
We propose a fuzzing framework that manages the complexities posed by fuzzing a diverse range of memory mapped peripherals without the need to generate register models. Existing approaches~\cite{p2im,uEmu} follow a method of first generating a model to define how memory mapped peripheral reads should be responded to. Other approaches such as Fuzzware~\cite{fuzzware} expand the list of fuzzed registers to include all peripheral registers and thus remove the need to determine the types of registers, such as \textit{Status} or \textit{Control}. Instead, Fuzzware uses symbolic execution to generate models that minimise the number of bytes required to be read from the fuzzer for each peripheral read. In contrast to generating models, we are motivated to investigate an alternative strategy for several reasons:

\begin{itemize}
    \item First, models based on heuristics are prone to failure, which can significantly reduce code coverage and prevent bugs from being triggered. Human analysis and intervention is needed for each firmware to prevent this from occurring.
    \item Second, modeling approaches declaring immutable inputs to registers could prevent testing of values that can be influenced by malicious actors. For example, bits within a serial port status register, such as the parity error, could be triggered by a malicious actor, resulting in the execution of the associated error handler. Notably, IFIZZ~\cite{ifizz}, previously identified error handlers as a common source of bugs, with more than 25\% of patches for some programs commonly used on routers containing changes to error handlers.
    \item Third, new models may need to be generated during fuzzing as the corresponding peripherals are discovered, complicating the fuzzing process. Each new model may retroactively impact the execution flow of previously discovered test cases. Consequently, the fuzzer may need to re-explore sections of previously explored firmware.
\end{itemize}

However, there are consequences to removing modeling methods. It reintroduces the problem of searching for valid values to overcome hurdles in firmware initialisation and reaching deep code paths as we discussed in Section~\ref{sec:problem}. In our study, we focus on the effectiveness of a fuzzer's mutation efforts on inputs likely to uncover interesting code-paths and we explore two observations to allow faster searching of the large input-space to discover valid, deep reaching inputs. 

\begin{itemize}
    \item \textbf{Observation 1: Register Value Behaviour in Firmware.~}We exploit the behaviour of register values characteristic of embedded peripheral registers using the proposed \textit{Peripheral Input Playback} technique. This technique provides mutations more likely to reach previously unseen code whilst recognising the need for registers to repeatedly report specific values in order to execute firmware correctly. We expand on our observation and our technique to exploit this behaviour in Section~\ref{sec:reg-val-behaviour}.
    \item \textbf{Observation 2: Uninformative Coverage Feedback.~}To address the underlying cause of the observation and to assist fast exploration of the input search space, we propose a new coverage instrumentation method, \textit{\fcov{}} described in Section~\ref{sec:duplicate-paths}. \fcov{} increases the likelihood of the mutation efforts on inputs to uncover interesting code-paths.
\end{itemize}

In the next Section, we describe and illustrate our observations, provide a conceptual view of the techniques we propose to utilize and defer the design and implementation details of our fuzzing framework to Section~\ref{sec:implementation}.

\subsection{Register Value Behaviour in Firmware}\label{sec:reg-val-behaviour}

Testing a register to infer the readiness of a peripheral device to accept or deliver new data is a very common occurrence during firmware execution. For example, according to an STM32 reference manual~\cite{stm32_reference}, a USART status register includes bits on whether the line is idle, read data is pending, the transmit buffer is empty, the last transfer is complete, and bits representing different errors. Often, such register values need to be repeatedly set to deliver data to the firmware to execute code that consumes the data. Hence, the desire is to primarily generate inputs that mimic typical use cases where data is available and no errors are reported to reach new or deeper code blocks. 

Consider the simplified example shown in Listing~\ref{fig:snippet_error_check}. To continue providing input to the firmware to execute code that processes the data, without repeatedly calling an error handler, the peripheral must have the \texttt{RX\_NOT\_EMPTY} bit set, with each of the error bits unset. While generating a single input that matches this pattern is trivial, repeatedly generating values following this pattern given the wide search space represented by all the possible register values is challenging. Consequently, it is more likely for a fuzzer to generate inputs to repeatedly trigger the \texttt{Error\_Handler} than those that result in the firmware receiving data (i.e executing \texttt{USART.Read()}). Thus, repeating an appropriate register value is necessary to provide deep code exploration.

\begin{listing}[h]
\begin{minted}{prolog}
Status = USART.SR 
if(Status & (FRAME_ERR | OVERRUN_ERR |
        NOISE_ERR | PARITY_ERR))
    Error_Handler(Status) 
else if(Status & RX_NOT_EMPTY)  
    return USART.Read() 
\end{minted}
    \caption{Simplified USART Error check code snippet}
    \label{fig:snippet_error_check}
\end{listing}

While the problem above demonstrates the need to repeat values to continue to provide new input to the firmware, it can also be necessary to succeed in the initialisation phase. Consider Listing~\ref{fig:snippet_status_check} that highlights an initialisation challenge encountered in the Drone firmware tested in our evaluation. This function waits for the peripheral to report it has finished transmitting a byte, then returns \texttt{OK}. If the correct bit is not set prior to a timeout, or no acknowledgement was reported by the peripheral, an error is returned. While setting a single bit is enough to overcome any returned errors, this function is called dozens of times during initialisation. Even a single reported failure during initialisation causes the firmware to abort. Thus, repeatedly reporting the \texttt{I2C\_TXEMPTY} bit as set is necessary to allow fuzzing of the full firmware.

\begin{listing}[h]
\begin{minted}{prolog}
Status i2c_wait_for_tx(device, timeout) 
    while(device.SR & I2C_TXEMPTY == 0) 
        if(device.SR & I2C_ACKFAILED) 
            return ERROR 
        if(getTime() > timeout) 
            return TIMEOUT 
    return OK 
\end{minted}
    \caption{Simplified I2C Status check code snippet based on a drone firmware~\cite{drone_i2c_status}}
    \label{fig:snippet_status_check}
\end{listing}

We recognize that prior work has also established the benefits associated with repeating values for reads to the same register. For control registers, \ptwoim{} returns the last value written by the firmware, and {\textmu}Emu does the same for \textit{Tier~0} registers. Additionally, a fixed constant value is returned for status registers in \ptwoim{} and \textit{Tier~1} registers in {\textmu}Emu.

However, we observe some cases where constantly repeating values is not desirable. Consider Listing~\ref{fig:snippet_reset_usart} showing an excerpt of code from the serial controller for a CNC firmware~\cite{cnc_snippet}. To proceed through a call to the \texttt{serial\_reset\_read\_buffer} function, the serial status register must report that no bytes are available to be read in the \texttt{uart\_tstc} function. However, the \texttt{uart\_tstc} function is queried to check for received data. Hence, to both initialise and take input from the serial port, \textit{status registers must have the ability to change}, even at the same point (or program counter) in the firmware. In addition to the ability to trigger error handlers, cases such as this \textit{motivates allowing the fuzzer to manipulate the values in status registers}.

\begin{listing}[h]
\begin{minted}{prolog}
int usart_tstc() 
    return (usart->SR & USART_SR_RXNE) != 0; 

void serial_reset_read_buffer() 
  while (usart_tstc()) 
    usart_getc()
\end{minted}
    \caption{Serial port reset code from CNC~\cite{cnc_snippet}}
    \label{fig:snippet_reset_usart}
\end{listing}

\subsection{Peripheral Input Playback Technique}
Based on the observation that repeated values are often needed to reach deep code paths, and the knowledge that generating many of these values is time consuming, we propose a method to define and inject value repetitions for a register.

In contrast to strictly employing the fuzzer generated input stream for register values, we use parts of the input stream to define a repetition scheme that influences the register behaviour.

The repetition scheme is defined based on subsections of the fuzz input, hence, random mutations by the fuzzer can still control \textit{the number of repetitions of a register value} to be arbitrarily large or small, provided the fuzzer generated input continues to improve coverage. Importantly, our approach allows for repeating of values as often as determined by the fuzzer without eliminating the potential to mutate values. Our approach aims to simplify the fuzzer's process of generating repeating sequences for a register, but \textit{does not prevent any possible sequences of values from being generated}.

This allows the sequence of status register values necessary to make progress through the code described in Listing~\ref{fig:snippet_reset_usart}. Additionally, by allowing the fuzzer to define repetitions of a register, valid values for the status registers in Listing~\ref{fig:snippet_error_check}~and~\ref{fig:snippet_status_check} only need to be solved once. Repetitions allow later accesses to continue to retrieve valid values, reducing the amount of fuzzer generated data required to overcome these hurdles which addresses the search space problem.

\subsection{Uninformative Coverage Feedback}\label{sec:duplicate-paths}

The number of paths identified by existing fuzzing frameworks is large compared to the size of the program. We observed path counts several times the block count reached in the program. This indicates that a large number of unique edges between different blocks have been reached. Given that function calls to predefined functions represent a single edge, and branch instructions have two potential edges, depending on whether the branch is taken or not, the excess of identified paths suggests \textit{coverage information is uninformative} and is being influenced by other aspects of the framework. 

One of the aspects unique to embedded software is the dependence on hardware interrupts. Hardware interrupts trigger an interrupt handler---generally a small piece of code that updates the state to influence later executions when the state is checked---then returns to the point in the program prior to the interrupt being triggered. The exact point in the program where the interrupt is triggered is not usually important; as the state modified by an interrupt is only checked periodically. Triggering the interrupt anywhere in this period will lead to the same outcome. Existing approaches such as \ptwoim{}, {\textmu}Emu and Fuzzware~\cite{fuzzware} periodically trigger interrupts in a round robin approach after a number of basic blocks have been executed. The point in the program where the interrupt is triggered can be influenced by the input, as the input affects the path taken, and by extension, the number of blocks executed before reaching a point in the program. Hence, the fuzzer will consider each unique point for triggering an interrupt a new interesting path, as it contains a new edge. 

\begin{figure}[h]
    \centering
    \includegraphics[width=0.75\linewidth]{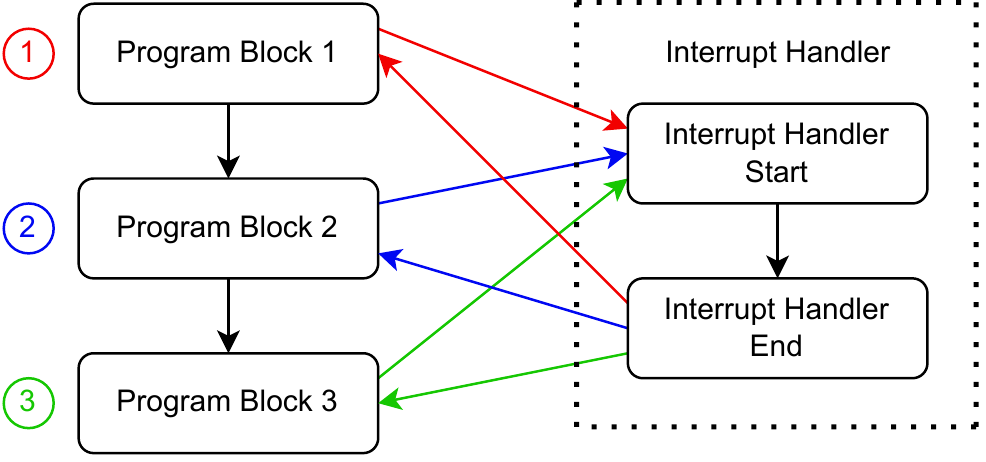}
    \caption{Jumps to and from an interrupt handler from each potential trigger point within a main program function. Each pair of edges jumping to, and returning from the interrupt handler marks a unique path from the fuzzer's view despite being functionally equivalent.}
    \label{fig:int_mask}
\end{figure}

Figure~\ref{fig:int_mask} illustrates an example with three potential pairs of edges caused by triggering an interrupt at varying points within a function to illustrate the problem. Depending on which of the three blocks within this function the interrupt is triggered at, a distinct new pair of edges will be created. For example, triggering an interrupt at \circled{1}, results in an edge from \textit{Program Block~1} to \textit{Interrupt Handler Start}, and a returning edge from \textit{Interrupt Handler End} back to \textit{Program Block~1}. Similarly, interrupts can be triggered at points \circled{2} and \circled{3}, with their corresponding edges now being present in the coverage information reported to the fuzzer. Consequently, it is feasible for three paths to be considered unique by the fuzzer, corresponding to the exact moment the interrupt is triggered. Given that the vast majority of blocks do not read the system state and will not be directly influenced by the interrupt, these paths are all \textit{functionally equivalent}.

We expect this problem to be most prevalent when loops are executed. In the case where the input defines the number of iterations of the loop, every value could potentially be considered a unique path, even if no new code is covered, as the point after the loop that the next interrupt occurs will change. In the case where the loop is a fixed number of iterations, by taking or not taking branches within the loop, the number of blocks executed in each iteration can be varied by the input, potentially leading to the same outcome.

A consequence of the uninformative coverage is poor scheduling of inputs (queue entries), as there are significantly more inputs to schedule, many of which are very similar or \textit{functionally equivalent}. Hence, time consumed applying mutations to these inputs is wasted if an equivalent entry has already been thoroughly explored.

\subsection{Functionally Equivalent Coverage Instrumentation (\fcov{}) Technique}

We propose changing the code coverage instrumentation employed by the emulator during fuzzing to remove functionally equivalent queue entries. \fcov{} separates the coverage instrumentation for interrupt control flow from the rest of the firmware control flow. The edges from both interrupt and non-interrupt coverage are then aggregated to a single coverage map. As a result, the context of an interrupt is obscured from the fuzzer's perspective. The coverage instrumentation technique prevents otherwise uninteresting inputs from being saved by the fuzzer. In the case where triggering the interrupt influences the system state, we expect that new code blocks would be executed, the corresponding edges would then ensure that the new path is saved and added to the queue. However, removing excess paths, we can more efficiently focus on paths that trigger new, deeper code.

\section{System Design and Implementation}\label{sec:implementation}
\begin{figure*}[!htbp]
    \centering
    \includegraphics[width=0.9\linewidth]{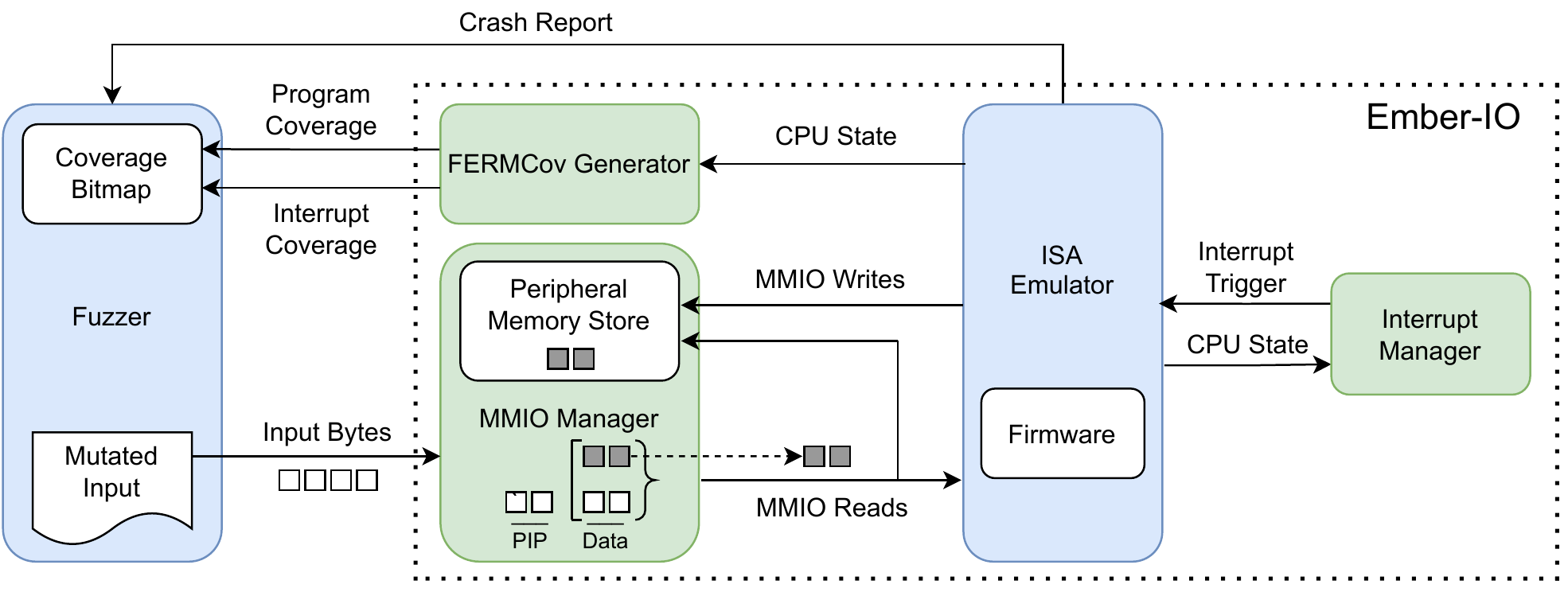}
    \caption{\emu{} design overview. Our implementation used AFL++~\cite{aflplusplus} for the Fuzzer, and QEMU~\cite{qemu} as the ISA emulator. The MMIO manager controls the raw input feed to the the Firmware requesting access to peripheral devices whilst the \fcov{} generator provides coverage feedback to the Fuzzer. The Peripheral Input Playback technique described in  Section~\ref{sec:input_playback} and Algorithm~\ref{alg:playback} is implemented with the MMIO Manager and Peripheral Memory Store. The firmware specific coverage instrumentation method described in Section~\ref{sec:interrupt_masking} and Algorithm~\ref{fig:interrupt_masking_code} is implemented with the \fcov{} Generator. The Interrupt Manager triggers an interrupt when a sleep instruction is encountered as described in Section~\ref{sec:implement_extras}, and periodically based on the number of blocks executed.}
    \label{fig:design_block}
\end{figure*}
Our solution is built on the QEMU mode of AFL++~\cite{aflplusplus}. We updated the AFL++ release of QEMU~\cite{qemu} to operate in system mode and function without variability in the presence of simulated hardware interrupts. A design overview of our implemented modules and their interactions are shown in Figure~\ref{fig:design_block}. We implement our techniques for the ARM architecture. Similar to \ptwoim~\cite{p2im}, we developed an interrupt manager to trigger interrupts using a round robin approach, with an interval defined by the number of basic blocks executed since the last interrupt. The set of enabled interrupts is gathered from the CPU state provided by the emulator. To handle some edge cases for execution, additional changes described in Section \ref{sec:implement_extras}, were made.

The execution of a test case is \textit{terminated} when one of four conditions is encountered:  
\begin{enumerate}
    \item A peripheral is read, and the input buffer is empty.
    \item Unmapped memory is accessed (Considered as crash).
    \item A memory permission error occurs, such as writing to read-only memory (Considered as crash).
    \item An instruction that jumps to itself is encountered (Considered as program termination).
\end{enumerate}

By default, memory permissions are configured to restrict the flash address space to read and execute, and RAM and peripheral memory to read and write. However, \emu{} can be configured to relax these permissions to support firmware that write to flash memory. The framework can also be configured to remove termination conditions 3 or 4. Condition 3 was disabled for testing with the \textit{3D Printer}, \textit{GPS Tracker}, \textit{uTasker MODBUS} and \textit{uTasker USB} firmware in our evaluation, as they write to flash memory located in the read only region. Condition 4 should be disabled for firmware with no main loop body, instead executing entirely inside of interrupt handlers. This was done for the \textit{Robot} and \textit{Soldering Iron} firmware in our evaluation.

\subsection{Peripheral Input Playback (MMIO Manager)}\label{sec:input_playback}

We implement the technique we refer to as peripheral input playback within the MMIO manager of our framework (Figure~\ref{fig:design_block}). We assign our own functions for handling reads and writes to the peripheral memory regions. Within these memory operation handlers, we implement the value repetition scheme described in Algorithm~\ref{alg:playback}. Prior to a register being read, a small value from the fuzzer generated input stream is read; we employ a two-bit value in our implementation. If this two-bit value is equal to 3 (\texttt{REPEAT\_CONST}), instead of responding with the next value in the input buffer, the previous value from this register is retrieved from the peripheral memory store and returned. The previous value represents either the value returned on the last read, or the value last written by the firmware---whichever was more recent. We aim to minimise the number of bits used for this purpose to reduce overhead. However, we do not want to excessively force repetition. The two-bit value comparison is trivially solvable by the fuzzer, and allows triggering only in the minority of cases, preventing excessive repetition.

\begin{algorithm}[t]
\begin{algorithmic}
\State \textbf{Input:} address
\State \textbf{Global:} periph\_memory\_store, REPEAT\_CONST
\State \textbf{Local:} register
\State \textbf{Output:} MMIO\_value
\State
\texttt{\textbackslash{}\textbackslash{}Check if we have previously\\\textbackslash{}\textbackslash{}accessed this register}
\If {address $\not\in$ periph\_memory\_store} 
    \State periph\_memory\_store $\gets$ \Call{NewRegister}{address}
\EndIf
\State register $\gets$ periph\_memory\_store[address]
\If{\textit{next value from fuzzer input} == REPEAT\_CONST}
\State\texttt{\textbackslash{}\textbackslash{}Repeat value from memory store}
\State MMIO\_value $\gets$ register.last\_value
\Else
\State\texttt{\textbackslash{}\textbackslash{}Update to new value}
    \State MMIO\_value $\gets$ \textit{next value from fuzzer input}
\EndIf
\State\texttt{\textbackslash{}\textbackslash{}Save value to memory store}
\State register.last\_value $\gets$ result
\State \Return MMIO\_value

\caption{Peripheral Input Playback Algorithm. Represents the memory read function for a peripheral address. The number of bits read from fuzzer input that is compared with REPEAT\_CONST is significantly smaller than the size of the register being read.}
\label{alg:playback}
\end{algorithmic}
\end{algorithm}

Given a random input test case, we expect this playback to be triggered on one in four register accesses when this implementation is used. We maintain a list of all registers encountered in the current test case, representing \texttt{periph\_memory\_store} in the given algorithm. In the case where a repetition is requested on the first access, we assume an initialisation value of 0. To maintain byte alignment on the input stream, and improve spatial locality of related bits, we read 32 bits from the input stream rather than 2 bits when determining if the value should be replayed. These 32 bits are used for the next 16 reads of that register. The improved spatial locality has a unique impact when considered in combination with the types of mutations a fuzzer can make. One mutation type in AFL based fuzzers is to replace an existing value with one of a set of predefined interesting values. Amongst these values are 0 and -1. When interpreted as binary data, the 32-bit value 0 would introduce no repetition and -1 would result in all of these reads playing back the previous value. For a firmware that repeatedly polls a 32-bit register for an appropriate value, only 8 bytes are required to report the correct value 16 times (4 bytes for the value, and 4 bytes for repetition information). Without this technique, 64 bytes of valid fuzzing input would have to be determined to overcome the same obstacle.

\subsection{\fcov{}}\label{sec:interrupt_masking}

Our implementation of \fcov{} modifies the existing code-coverage instrumentation in AFL++'s QEMU mode based on the CPU state. As shown in Algorithm~\ref{fig:interrupt_masking_code}, any jumps to an interrupt handler have the \textit{last\_int\_block} variable set to null during the previously executed program block. Therefore, from the perspective of the fuzzer, all jumps to an interrupt handler are seen as from the same location. The return jump when exiting the interrupt handler is entirely excluded from the coverage map. By keeping the value of \textit{last\_program\_block} untouched during the interrupt, continuity of the observed non-interrupt edges can be maintained.

\begin{algorithm}[t]
\begin{algorithmic}
\State \textbf{Input:} cpu
\State \textbf{Global:} last\_program\_block
\State \textbf{Global:} last\_int\_block
\State \textbf{Global:} coverage
\State \textbf{Local:} current\_block
\State current\_block $\gets$ cpu.program\_counter
\If{\textit{cpu in interrupt}}
    \State\texttt{\textbackslash{}\textbackslash{}Add interrupt edge to coverage map}
    \State coverage $\gets$ \Call{add\_edge}{last\_int\_block, current\_block}
    \State last\_int\_block $\gets$ current\_block
\Else
    \State\texttt{\textbackslash{}\textbackslash{}Add program edge to coverage map}
    \State coverage $\gets$ \Call{add\_edge}{last\_program\_block, current\_block}
    \State last\_program\_block $\gets$ current\_block
    \State last\_int\_block $\gets$ $null$
\EndIf
\State \Return
\caption{\fcov{} Algorithm. Represents a function called when each block is executed}
\label{fig:interrupt_masking_code}
\end{algorithmic}
\end{algorithm}

We determine the interrupt state by probing QEMU's emulated Nested Vectored Interrupt Controller (NVIC). This implementation does not require any comparison between different paths in order to remove functionally equivalent paths, instead relying on the coverage map checks already present at the end of each executed test case, minimising processing overhead.

\subsection{Additional Components}\label{sec:implement_extras}
To address a couple of firmware specific problems, we make some additional extensions to our implementation. The first of these problems is the use of power saving features. Firmware may put the CPU to sleep, and wait for a hardware interrupt to occur before proceeding. In emulators such as QEMU that implement these instructions, this can be problematic. Our interrupt manager is based on the design used in \ptwoim{}~\cite{p2im}, that periodically triggers interrupts based on the number of blocks executed by the emulator. But if the CPU is sleeping, no further blocks will be executed, thus no interrupt will be triggered to wake the CPU. We extend the interrupt manager to trigger an interrupt when the CPU enters sleep mode.

Additionally, we observed some firmware with extremely long initialisation routines, far exceeding the time in the main program loop. To ensure performance is not compromised in this case, we allow the fuzzer to operate with multiple forkservers operating at varied depths in the firmware under test. A random forkserver is chosen before each execution, allowing some tests to run from after the initialisation stage is complete.

\section{Evaluation}

We design our experimental regime to answer the following four key questions:

\begin{enumerate}
    \item What is the impact of the techniques employed by \emu?
    (Section~\ref{sec:impact})
    \item Is \emu{'s} model-free approach applicable to a wide range of firmware and hardware platforms?
    (Section~\ref{sec:perf-comp})
    \item How does \emu{'s} performance compare with previous \textit{re-hosting} methods developed for fuzzing monolithic firmware? (Section~\ref{sec:coverage})
    \item Can \emu{} be used to discover previously unknown software bugs in real world firmware? (Section~\ref{sec:crashes})
\end{enumerate}
     
\noindent\textbf{Settings.~}Unless otherwise specified: all \emu{} experiments were carried out using a modified copy of AFL++ 3.15a as the fuzzer. All AFL++ settings were kept as default, except to adjust the timeout for hang detection when necessary. We describe any evaluation specific settings the start of the relevant sections below. 


\subsection{Impact of Techniques}\label{sec:impact}
To verify the usefulness of the techniques we explore, we take a sample of existing firmware from previous works and introduce one technique at a time to measure the improvement in block coverage. We selected the Drone and 3D printer firmware based on past performance in existing firmware fuzzing frameworks~\cite{p2im,uEmu,fuzzware}. The Drone binary provided some of the highest coverage in each of the existing frameworks, while the 3D Printer binary consistently performed poorly. By extension, we expect the peripheral accesses within these firmware to represent a mix between types already well recognised and dealt with in previous fuzzing approaches, and those that pose a challenge to these approaches.

The configurations are as set out as described at the beginning of Section~\ref{sec:implementation}. The \textit{Baseline} represents fuzzing campaigns found with our AFL++~QEMU mode modified to allow embedded system fuzzing, without additional techniques applied. Next, we introduce our peripheral input playback technique. Subsequently, we combine peripheral input playback with \fcov{}. Each test run was executed for 24 hours. Five runs were completed for each test.

\begin{table}[h]
\centering
\caption{Evaluating the impact of the proposed techniques in \emu. Median blocks covered and paths found with sequential introduction of techniques for Drone (from \ptwoim~\cite{p2im}) and 3D Printer (from \textmu{Emu}~\cite{uEmu}) binaries.}
\label{fig:individual_feature_comparison}
\resizebox{7.5cm}{!}{
\begin{tabular}{|l|l|r|r|} 
\hline
\begin{tabular}[c]{@{}l@{}}Firmware \end{tabular}                                                  & Techniques                                                                          & \multicolumn{1}{l|}{\begin{tabular}[c]{@{}l@{}}Blocks \\Covered\end{tabular}} & \multicolumn{1}{l|}{Paths}  \\ 
\hhline{|====|}
\multirow{3}{*}{\begin{tabular}[c]{@{}l@{}}Drone\end{tabular}}       & Baseline                                                                            & 1601                                                                          & 8759                       \\ 
\cline{2-4}
                                                                                                        & Peripheral Input Playback                                                                 & 1810                                                                          & 11737                       \\ 
\cline{2-4}
                                                                                                        & \begin{tabular}[c]{@{}l@{}}Peripheral Input Playback \\and \fcov{}\end{tabular} & 1836                                                                          & 1422                        \\ 
\hline
\multirow{3}{*}{\begin{tabular}[c]{@{}l@{}}3D Printer\end{tabular}} & Baseline                                                                            & 1594                                                                          & 5800                       \\ 
\cline{2-4}
                                                                                                        & Peripheral Input Playback                                                                 & 2114                                                                          & 8466                       \\ 
\cline{2-4}
                                                                                                        & \begin{tabular}[c]{@{}l@{}}Peripheral Input Playback\\and \fcov{}\end{tabular}  & 3336                                                                          & 1879                        \\
\hline
\end{tabular}
}
\end{table}

\subsubsection{Drone Firmware}
The drone firmware sets up multiple different peripherals, such as an IMU (inertial measurement unit), compass, barometer and telemetry module. After the peripherals have been successfully initialised, an advanced control algorithm is used to interpret the inputs and maintain the stability of the drone.

\begin{figure}[h]
    \centering
    \includegraphics[width=1.0\linewidth,trim={0 0.3cm 0.5cm 0cm},clip]{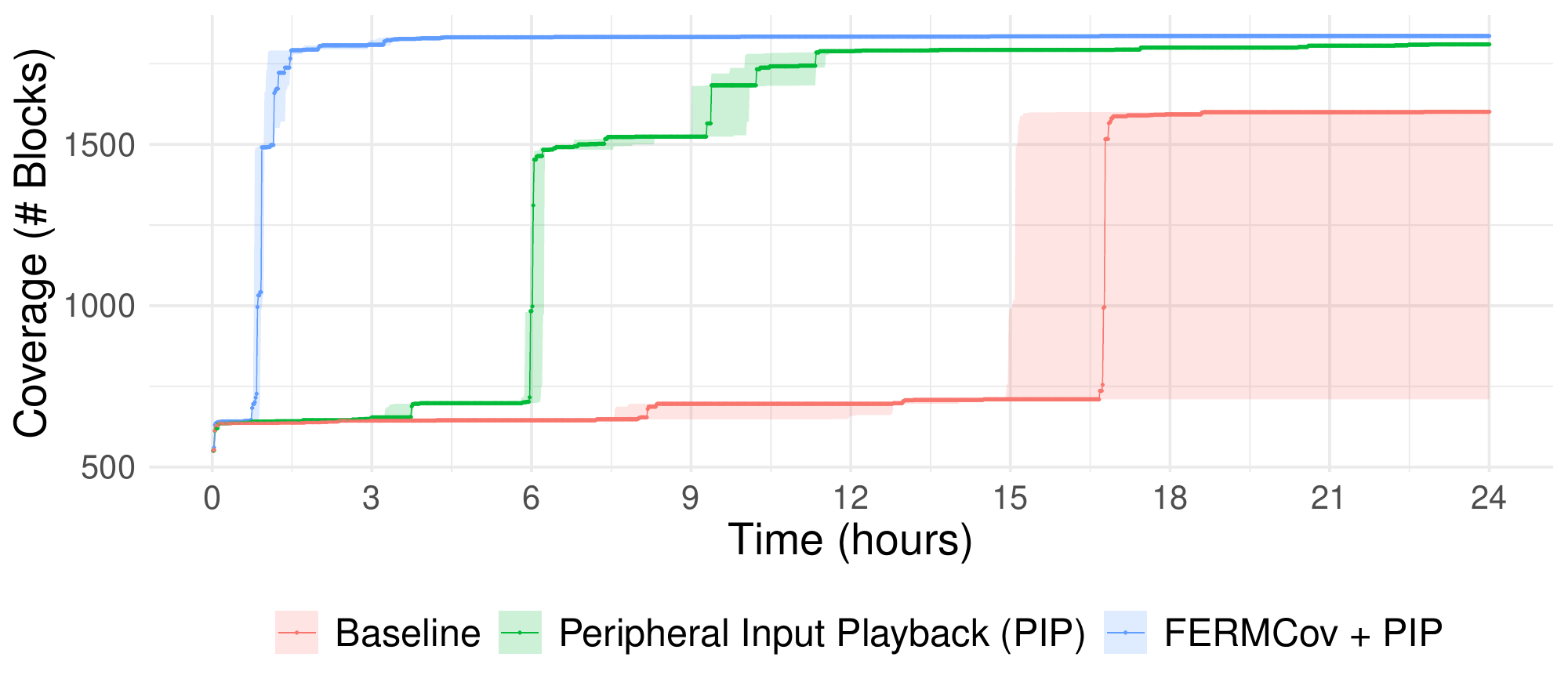}
    \caption{Drone firmware coverage band graph. Shaded band represents the coverage achieved by the middle 60\% of executions. Existing fuzzing frameworks reported high coverage in this binary. \textit{\textbf{Baseline}} represents fuzzing campaigns with our AFL++ QEMU mode modified to allow embedded system fuzzing, with no additional techniques applied.}
    \label{fig:drone_cov}
\end{figure}

Coverage over time for this firmware is shown in Figure~\ref{fig:drone_cov}. Peripheral input playback resulted in a 13\% increase in coverage in this binary. Additionally, the time required to complete the initialisation and reach the main program was reduced from more than 16 hours to 6 hours in the median case. The addition of \fcov{} reduced this time to less than one hour when compared to just using peripheral input playback.

We credit this speedup to the removal of functionally equivalent paths. 87.9\% \textit{fewer paths} were present in the fuzzing queue when \fcov{} was enabled, with a slight increase in blocks covered. The introduction of these techniques increased the median baseline block coverage result by 14.7\% compared to the baseline.

\subsubsection{3D Printer Firmware}

\begin{figure}[h]
    \centering
    \includegraphics[width=1.0\linewidth,trim={0 0.3cm 0.5cm 0}, clip]{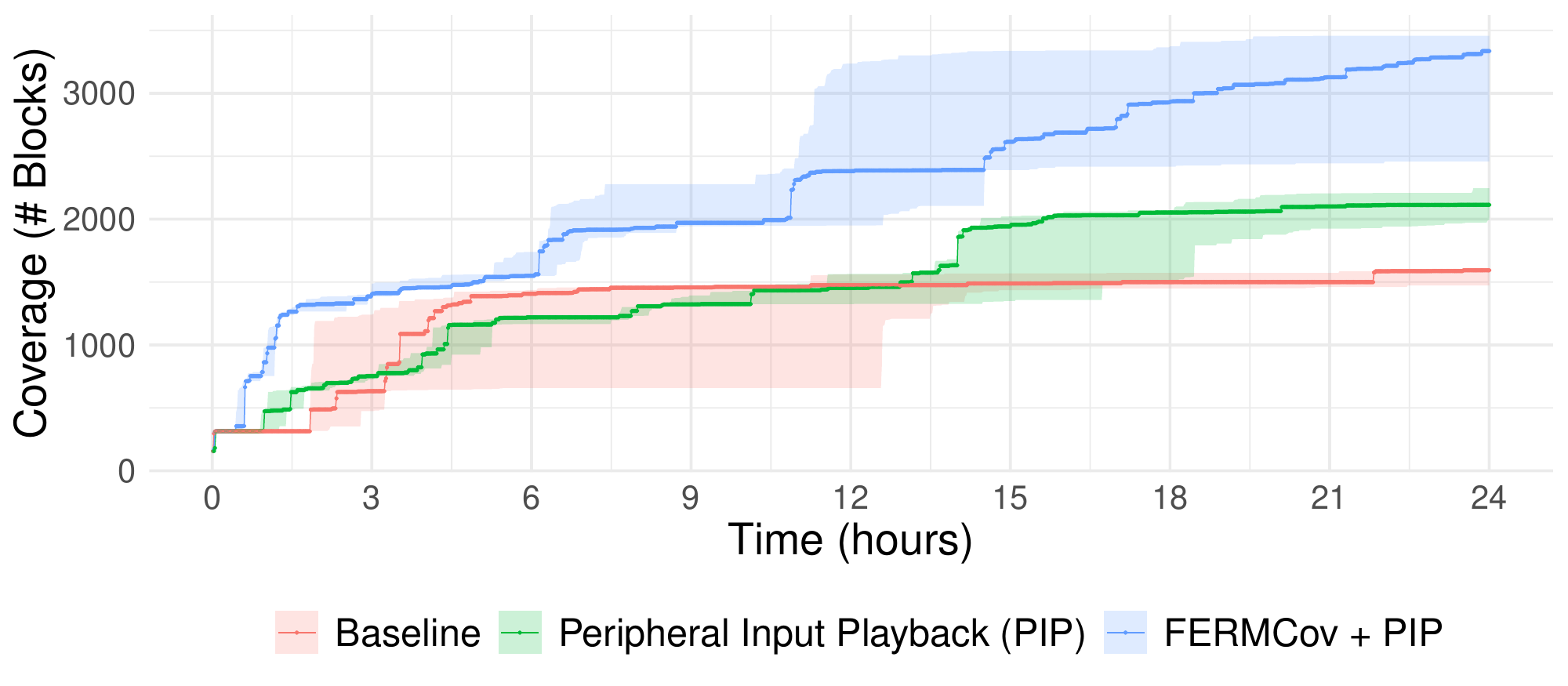}
    \caption{3D Printer firmware coverage band graph. Shaded band represents the coverage achieved by the middle 60\% of executions. Existing fuzzing frameworks report low coverage in this binary. \textit{\textbf{Baseline}} represents campaigns with our AFL++~QEMU mode modified to allow embedded system fuzzing, with no additional techniques applied.} 
    \label{fig:3dp_cov}
\end{figure}

Peripheral input playback provided a significant improvement in the second half of the fuzzing session, as shown in Figure~\ref{fig:3dp_cov}, with more than a 30\% increase in blocks covered after 24 hours. The addition of \fcov{} provided further increased coverage, 109.3\% ahead of the baseline after 24 hours. The number of paths in the queue with peripheral input playback and \fcov{} enabled was 77.8\% lower than when peripheral input playback was used on its own. The results confirm our expectations and demonstrate the effectiveness of the approach we propose.

\subsection{Comparison with State of the Art}\label{sec:perf-comp}

We consider the existing state-of-the-art emulation based embedded fuzzing framework {\sc Fuzzware}~\cite{fuzzware} for our comparison. Five trials are performed for each firmware, which are executed for 24~hours as in previous works. Our tests were conducted using an AMD Threadripper 3990X and 256GB of RAM. We employed the suite of binaries used to compare coverage in \textmu{Emu}~\cite{uEmu} and Fuzzware~\cite{fuzzware} to generate our experimental results. 

In addition to running {\sc Fuzzware}~\cite{fuzzware} as released with default settings, we extend {\sc Fuzzware} with our \fcov{} technique to remove functionally equivalent paths. We do not implement peripheral input playback as it conflicts with Fuzzware's techniques, which only populates a subset of the bits in each register depending on the model for that register and program counter (PC) pair.

For the firmware that interacts with a DMA enabled peripheral, we manually configured the DMA related registers as passthrough following {\sc Fuzzware}. The XML Parser binary required the disabling of interrupts in the emulator, as configured by prior works. All other firmware was successfully run without modification to the firmware or manual configuration of peripherals.

\subsubsection{Coverage}\label{sec:coverage}

The number of blocks covered for each fuzzer and binary is presented in Table~\ref{fig:pval_tab}. To quantify the significance of the differences in code coverage between \emu{} and Fuzzware, we adopt the testing methodology previously used by Fuzzware~\cite{fuzzware}. Statistical significance is determined from P-values calculated using Mann Whitney U Tests, at a 0.01 significance level. Table~\ref{fig:pval_tab} summarises the results of the evaluation campaign. In ten of these firmware, no statistically significant difference was observed when compared to Fuzzware's model-based approach. However, of the eleven firmware where a statistically significant difference was observed, \emu{} achieved higher coverage in \textit{eight}, while the remaining three firmware performed better under {\sc Fuzzware}. 

The 3D Printer, LiteOS IoT and Thermostat firmware all saw a median coverage increase exceeding 10\%, achieving a 254.6\%, 11.1\% and 32.8\% increase respectively. The 6LoWPAN Receiver, 6LoWPAN Sender, Steering Control and uTasker MODBUS binaries each show a moderate improvement in median coverage between 5 and 10\%. For the Robot binary we observe a consistent, but very small increase in coverage of less than 1\%. The CNC and Soldering Iron firmware both favoured {\sc Fuzzware} with an increase of 10.9\%, while the GPS Tracker firmware favoured {\sc Fuzzware} by 2.5\%. Notably, despite the lower coverage results in these binaries, \emu{} reproduced the bugs found using {\sc Fuzzware}. Coverage graphs with the minimum, median and maximum values for each binary and fuzzing framework are shown in Figure~\ref{fig:cov_plot} in \textbf{\textit{Apendix~\ref{apd:coverage-comp}}} while investigations into increased coverage is discussed in \textbf{\textit{Apendix~\ref{apd:coverage-analysis}}}.

Applying our \fcov{} technique to {\sc Fuzzware} yielded statistically significant differences in the 3D Printer and uTasker MODBUS binaries. In both cases \fcov{} provided an increase in code coverage; 31.7\% for the 3D Printer and 5.9\% for the uTasker binary. No statistically significant negative impacts were observed from the introduction of this technique into {\sc Fuzzware}. 

\begin{table*}
\centering
\caption{Comparing \emu{} with the state-of-the-art--{\sc Fuzzware}. We report the minimum, median and maximum block coverage achieved with each fuzzing framework over five 24 hour fuzzing campaigns. P-Values indicate statistical significance (calculated using Mann Whitney U tests, conducted at a 0.01 significance level) between: i)~{\sc Fuzzware} and \emu{}; and ii)~{\sc Fuzzware} and {\sc Fuzzware} with \fcov{}. The highest minimum, median and maximum values for each firmware are shown in \textbf{bold}.}
\label{fig:pval_tab}
\resizebox{\linewidth}{!}{
\begin{tabular}{|l|l||l|l|l||l|l|l|l||l|l|l|l|} 
\cline{3-13}
\multicolumn{1}{c}{} & \multicolumn{1}{c|}{}                                          & \multicolumn{3}{c|}{Fuzzware}                 & \multicolumn{4}{c|}{Fuzzware with \fcov{}}             & \multicolumn{4}{c|}{Ember-IO}                            \\ 
\hline
Firmware             & \begin{tabular}[c]{@{}l@{}}\# Blocks in\\Firmware\end{tabular} & Min           & Median        & Max           & Min           & Median        & Max           & p-value & Min           & Median        & Max           & p-value  \\ 
\hline
3D Printer           & 8045                                                           & 851           & 919           & 937           & 1187          & 1210          & 1333          & <0.01    & \textbf{2548} & \textbf{3259} & \textbf{3667} & <0.01     \\ 
\hline
6LoWPAN Receiver     & 6977                                                           & 3016          & 3057          & 3164          & 2979          & 3144          & 3182          & 0.347   & \textbf{3255} & \textbf{3284} & \textbf{3295} & <0.01     \\ 
\hline
6LoWPAN Sender       & 6980                                                           & 2995          & 3009          & 3147          & 3029          & 3139          & 3149          & 0.047   & \textbf{3268} & \textbf{3289} & \textbf{3301} & <0.01     \\ 
\hline
LiteOS IoT           & 2423                                                           & 738           & 1213          & 1339          & 1205          & 1339          & 1347          & 0.141   & \textbf{1347} & \textbf{1348} & \textbf{1350} & <0.01     \\ 
\hline
Robot                & 3034                                                           & 1289          & 1303          & 1307          & 1299          & 1305          & 1308          & 0.389   & \textbf{1308} & \textbf{1315} & \textbf{1323} & <0.01     \\ 
\hline
Steering Control     & 1835                                                           & 609           & 611           & 615           & 609           & 613           & 615           & 0.390   & \textbf{638}  & \textbf{644}  & \textbf{648}  & <0.01     \\ 
\hline
Thermostat           & 4673                                                           & 3065          & 3227          & 3515          & 2971          & 3330          & 3470          & 0.754   & \textbf{4230} & \textbf{4285} & \textbf{4365} & <0.01     \\ 
\hline
uTasker MODBUS       & 3780                                                           & 1176          & 1176          & 1211          & 1217          & 1246          & 1285          & <0.01    & \textbf{1219} & \textbf{1252} & \textbf{1311} & <0.01     \\ 
\hline
Console              & 2251                                                           & 803           & 805           & 844           & \textbf{805}  & \textbf{844}  & 844           & 0.065   & 804           & 843           & \textbf{856}  & 0.168    \\ 
\hline
Drone                & 2728                                                           & 1830          & \textbf{1843} & \textbf{1845} & 1832          & 1841          & \textbf{1845} & 0.675   & \textbf{1834} & 1835          & 1839          & 0.173    \\ 
\hline
Gateway              & 4921                                                           & 2441          & 2848          & \textbf{2962} & \textbf{2569} & \textbf{2859} & 2938          & 0.917   & 2104          & 2214          & 2571          & 0.016    \\ 
\hline
Heat Press           & 1837                                                           & 534           & 540           & 553           & 531           & 538           & 545           & 0.600   & \textbf{549}  & \textbf{554}  & \textbf{554}  & 0.026    \\ 
\hline
PLC                  & 2303                                                           & 627           & 638           & 643           & \textbf{636}  & 640           & 665           & 0.209   & \textbf{636}  & \textbf{642}  & \textbf{810}  & 0.117    \\ 
\hline
Reflow Oven          & 2947                                                           & 1191          & \textbf{1192} & 1193          & \textbf{1192} & \textbf{1192} & 1192          & 0.519   & \textbf{1192} & \textbf{1192} & \textbf{1196} & 0.145    \\ 
\hline
RF Door Lock         & 3320                                                           & 781           & 782           & 2484          & 781           & \textbf{2282} & \textbf{2692} & 0.341   & \textbf{782}  & 782           & 2662          & 0.576    \\ 
\hline
uTasker USB          & 3491                                                           & 1360          & 1672          & 1730          & \textbf{1380} & \textbf{1681} & \textbf{1829} & 0.602   & 1319          & 1336          & 1393          & 0.016    \\ 
\hline
XML Parser           & 9376                                                           & 3635          & 3878          & 3953          & \textbf{3651} & \textbf{3884} & \textbf{4086} & 0.465   & 3449          & 3629          & 3832          & 0.047    \\ 
\hline
Zephyr SocketCan     & 5943                                                           & 2509          & \textbf{2714} & \textbf{2794} & \textbf{2517} & 2703          & 2768          & 0.834   & 2272          & 2468          & 2565          & 0.076    \\ 
\hline
CNC                  & 3614                                                           & 2526          & 2588          & 2700          & \textbf{2638} & \textbf{2671} & \textbf{2732} & 0.175   & 1781          & 2305          & 2447          & <0.01     \\ 
\hline
GPS Tracker          & 4194                                                           & \textbf{1010} & 1027          & 1036          & \textbf{1010} & \textbf{1032} & \textbf{1037} & 0.292   & 973           & 1001          & 1004          & <0.01     \\ 
\hline
Soldering Iron       & 3656                                                           & \textbf{2470} & 2541          & 2554          & 2315          & \textbf{2568} & \textbf{2580} & 0.465   & 2206          & 2265          & 2298          & <0.01     \\
\hline
\end{tabular}
}
\end{table*}

\subsubsection{Detection of Bugs}\label{sec:crashes}
The authors of Fuzzware~\cite{fuzzware} have created a list of bugs identified in firmware. From these binaries, they identified 35 bugs, of which one is a false positive. Using \emu{}, we have reproduced 32 of these known bugs, including the false positive. We further investigated the bugs that were not reproduced by our fuzzer. The first of these bugs (within the GPS Tracker binary) is guarded by a string comparison, and the authors of Fuzzware note that triggering this bug is highly non-deterministic~\cite{fuzzware_gps_bug}. The other two bugs (within the uTasker MODBUS binary) are dependent on interrupts being triggered while a variable is uninitialised. In our tests with \emu, we did not observe an instance of the required interrupts being triggered while the firmware was in the required state. However, we were able to reproduce the third bug in this binary, which is based on the same root cause, where the \textit{same variable} is used while uninitialised, in a different location. In addition to the known bugs, we observed crashes not attributable to these previously known bugs. The results of our analysis on these crashes are described below. 

\vspace{2mm}
\noindent\textbf{(1)~CNC.~}In addition to the known bugs in this binary, we identified a crash corresponding to a race condition. In rare cases, a function would be called that first verifies a pointer stored in a global variable is non-null before proceeding. However, the pointer can be modified inside of an interrupt handler. If an interrupt is triggered after the null check but \textit{before} the pointer is used, the interrupt handler can modify the pointer, including setting it to null. After further analysis we consider this to be a false positive, as on the real-hardware, the interrupt will not be triggered while in the prerequisite state.

\vspace{2mm}
\noindent\textbf{(2)~Drone.~}The identified crashes in this firmware are caused by an underflow because of unexpected interrupt timing. This underflow influences the size of the bytes remaining in a UART transfer. The large value for the transfer size causes the firmware to read from unmapped memory. We consider this bug a false positive, as it requires a control register value to change during execution, without the firmware writing to it, which would not occur for this register on real hardware.

\vspace{2mm}
\noindent\textbf{(3)~Gateway.~}When processing system messages, the length of the message is not properly validated. As data is received, each byte is buffered, until the message is terminated. While the firmware performs a bounds check before storing the received byte into the buffer, a separate variable that keeps track of the number of received bytes is incremented even when the byte is not stored. Corruption then occurs when the message is parsed with a size greater than the buffer holds. We consider this to be a security bug.

\vspace{2mm}
\noindent\textbf{(4)~Zephyr SocketCan.~}We observed crashes based on a stack buffer overflow. During the parsing of text commands in the console, the number of arguments is not correctly validated. This leads to the array terminator overwriting an adjacent saved register on the stack. This saved register corresponds to a pointer that is later popped from the stack and dereferenced. This bug has not been reported by existing firmware fuzzing frameworks, but was found and fixed by the software developers in a later version~\cite{zephyr_patch} prior to our discovery.

\vspace{2mm}
\noindent\textbf{(5 \& 6)~6LoWPAN Sender \& Receiver.~}We observed identical crashes in these two binaries based on the usage of an uninitialised variable. During firmware initialisation, there is a brief period after the interrupts are enabled for the serial port while a global pointer is uninitialised. By triggering an interrupt during this period, a null pointer is dereferenced within the interrupt, causing a crash.

\vspace{2mm}
\noindent
Given the knowledge of these new bugs, we investigated the crashes reported in these binaries by Fuzzware with \fcov{} applied. We were able to reproduce the observed crashes for the CNC, Drone and Zephyr SocketCan.

Additionally, we investigated the binaries where \emu{} did not identify all of the known bugs. In 24 hours, {\sc Fuzzware} with \fcov{} was able to reproduce the bug missed by \emu{} in the GPS Tracker, and one of the two bugs missed in the uTasker MODBUS binary. We note that running {\sc Fuzzware} alone did not yield crashes corresponding to the missed GPS Tracker bug, or either of the missed crashes in the uTasker MODBUS binary within 24 hours for any of the 5 trials. \textit{We provide a crash analysis of replicated and newly discovered bugs in Table~\ref{fig:crash_tab} in \textbf{Appendix~\ref{apd:crash-causes}}}.

\section{Discussion}\label{sec:limitations}
In this section, we discuss false positives, performance differences we observed between {\sc Fuzzware} and \emu{}, the applicability of \emu{} beyond the prototype implementation.

\vspace{1mm}
\noindent\textbf{Direct Memory Access.~}We demonstrate that \emu{} is effective through coverage performance analysis and discovery of new bugs using a suite of 21 firmware tested by existing firmware fuzzing frameworks. Notably, we have not considered firmware dependent on DMA and implemented techniques to automatically handle accesses to DMA. Existing techniques such as DICE~\cite{dice} can be applied along with \emu{} to fuzz firmware with DMA.

\vspace{1mm}
\noindent\textbf{False Positive Analysis.~}Although we have discovered six software bugs not reported by existing firmware fuzzing frameworks, it is important to recognise that the fuzzer generated values returned to memory reads do not follow constraints that may be present on physical devices. By assuming that any value can be contained in a register, paths or crashes may be encountered that are not reachable when the firmware is executed on the physical device in a deployment setting. We believe this is the case in two of the six bugs found. As a result, we do not expect to trigger these bugs on the original hardware, unless the peripheral malfunctions.

False positives are an inherent problem in fuzzing of device peripherals, as unknown hardware defines the set of possible inputs for a given register, which cannot be directly inferred from the firmware. Notably, this is less likely to occur in approaches such as \ptwoim~\cite{p2im} and \textmu{Emu}~\cite{uEmu} due to their enforcement of register models based on common access patterns. Some classifications, such as control registers for \ptwoim{} and T0 registers for \textmu{Emu} can only contain values written by the firmware, preventing unexpected inputs. However, fixed values and fuzzer inputs are not guaranteed to be values observable on physical devices. A concurrent work~\cite{semu} has looked at avoiding this issue using information from MCU reference manuals. Both \emu{} and Fuzzware consider all registers as valid sources to inject fuzzing input.

\vspace{1mm}
\noindent\textbf{Performance.~}We find the difference in performance between \emu{} and Fuzzware interesting in certain binaries. Of particular interest is the increase in performance of \emu{} for the 3D Printer binary, especially when compared to the CNC binary, a firmware that also interprets G-Code commands. We investigated these binaries further to identify the cause of the disparity. We observed a large number of calls to functions in the initialisation of the 3D Printer binary that write long text strings to a serial port. In order to return from each of these function calls, suitable status values need to be supplied for each byte to be written from the string. As the write functions are called more often, coverage feedback for the edges within this function begin to saturate, and intermediate steps towards uncovering the next block after the serial write become more distant. The use of peripheral input playback reduces the number of bytes required in a mutated input bridge the gap. While \fcov{} provides clearer feedback to the fuzzer, allowing the inputs that reach furthest through the initialisation phase to be more easily identified for further mutation. Both these techniques assist in completing steps in initialisation required to reach the main program loop.

Investigating the CNC binary further, we observe Fuzzware employs the use of its pass-through model extensively in this binary. This model is applied when the register value has no influence on the program state, and prevents any fuzzing input being used for these register reads. Thus, the number of bytes in the test case for a given path is reduced, improving the chance of a mutation uncovering a new edge. While our peripheral input playback technique allows for simpler solving of sequences of register accesses, our techniques do not target improvements in handling registers that cannot influence execution. We observed similar patterns in the Soldering Iron binary. This likely contributes significantly to the positive results obtained by Fuzzware in these binaries.

\vspace{1mm}
\noindent\textbf{Nested Interrupts.~}Nested interrupts are not triggered by either \emu{} or {\sc Fuzzware}. Hence, our current \fcov{} implementation does not attempt to resolve functionally equivalent paths caused by nested interrupts. However, \fcov{} could be adapted to separate the coverage for each interrupt, rather than interrupt code and program code to support removal of functionally equivalent paths caused by nested interrupts.

\vspace{-3mm}
\section{Related Work}

There have been many contributions to the area of embedded system testing. Many previous approaches~\cite{avatar2,conware,inception,embedded2016,avatar,pretender,unicorefuzz} base their execution on the observed execution of the hardware. These observations are then used to build an approximate model that can be used to emulate peripheral devices.

Targeted approaches to analyse specific, complex peripherals such as Bluetooth and Cellular baseband were previously investigated in a security context~\cite{firmxray,lightblue,basesafe}. While HALucinator~\cite{HALucinator} aims to minimize the effort to emulate a wide range of peripherals by performing this emulation at the Hardware Abstraction Layer (HAL). Hence, by targeting all devices that use the same HAL, the engineering effort is applicable to a wider variety of devices. However, this requires the target firmware to use a supported HAL.

Methods applied in FIE~\cite{FIE} for MSP430 microcontrollers look at symbolically executing every possible state in smaller programs. To deal with the state explosion, the authors optimized their implementation to remove paths that reach identical states. Laelaps~\cite{laelaps} uses symbolic execution to explore firmware. After identifying a local ideal path based on a set of heuristics, a concrete value is determined that allows execution to follow the identified ideal path. However, the maximum depth of execution before concretising is limited, therefore a globally ideal path cannot always be determined. Other recent works such as Jetset~\cite{jetset} and Mousse~\cite{mousse} have further explored symbolic execution in these environments.

Prior embedded system fuzzing frameworks such as \ptwoim~\cite{p2im} and {\textmu}Emu~\cite{uEmu} do not require manual implementation of any peripherals, nor do they require physical hardware to develop a model. \ptwoim{} uses heuristics based on the access patterns of registers to characterise their type as either status, control or data. The authors implement corresponding expected behaviours for each type. Input from the fuzzer only influences the values of data registers. By comparison, {\textmu}Emu uses invalid states as a means to determine an assumed register type was incorrect and update it until valid execution is achieved. This results in a knowledge-base presented as tiers of register types. A heuristic approach is then used to identify which registers to apply fuzzer input to.

The state-of-the-art firmware fuzzing framework Fuzzware~\cite{fuzzware} avoids assigning models that categorise registers into types. Instead it uses dynamic symbolic execution to generate a model that constrains register values to those that influence program logic by removing redundant bits from the input. By not assigning an expected behaviour to each register, all register accesses that influence execution are fuzzed, and any feasible program path can be explored.

The recent study, ICICLE~\cite{icicle} provides a mechanism to both execute and instrument firmware binaries in an architecture agnostic fashion, reducing the manual work required to fuzz firmware on new targets.

In general, many techniques exist to improve the mutation, scheduling or instrumentation of greybox fuzzers for desktop applications~\cite{redqueen, ijon, greyone, entropic, aflfast, mopt}. These approaches could be adapted to improve embedded systems fuzzing. 

\section{Conclusion}
To further capabilities of embedded firmware fuzzers, we developed \emu, a novel embedded system fuzzing framework for memory mapped IO without peripheral modelling. Replaying values allows the fuzzer to overcome repeated status register checks and significantly reduced the time required to overcome firmware initialisation. This allowed reaching code that baseline tests could not consistently reach in a reasonable time frame. By changing coverage instrumentation for firmware targets, we found, upwards of 75\% of paths could be removed as they were functionally equivalent to existing paths. Our testing found these paths can  be discarded without any reduction in code covered. The increased efficiency of the fuzzer with our proposed techniques reduced the time to overcome obstacles to firmware fuzzing and led to further improvements in coverage for a given time frame. Our model free approach improved code coverage by more than 250\% in the best case when compared to the state-of-the-art firmware fuzzer. Six firmware bugs were triggered that previous works had not identified, where two were triaged to be false positives. Notably, {\sc Fuzzware} with \fcov{} implementation improved the consistency of bug discovery (see Appendix~\ref{apd:crash-causes}) and coverage compared to employing {\sc Fuzzware}.

\section*{Acknowledgements}

This work was supported through the Australian Government's Research Training Program Scholarship (RTPS).

\bibliographystyle{plain}
\bibliography{references.bib}

\clearpage
\appendix
\section{Appendix}

\subsection{Coverage Over Time Comparisons}\label{apd:coverage-comp}

\begin{figure*}[bh]
    \centering
    \includegraphics[width=0.86\linewidth]{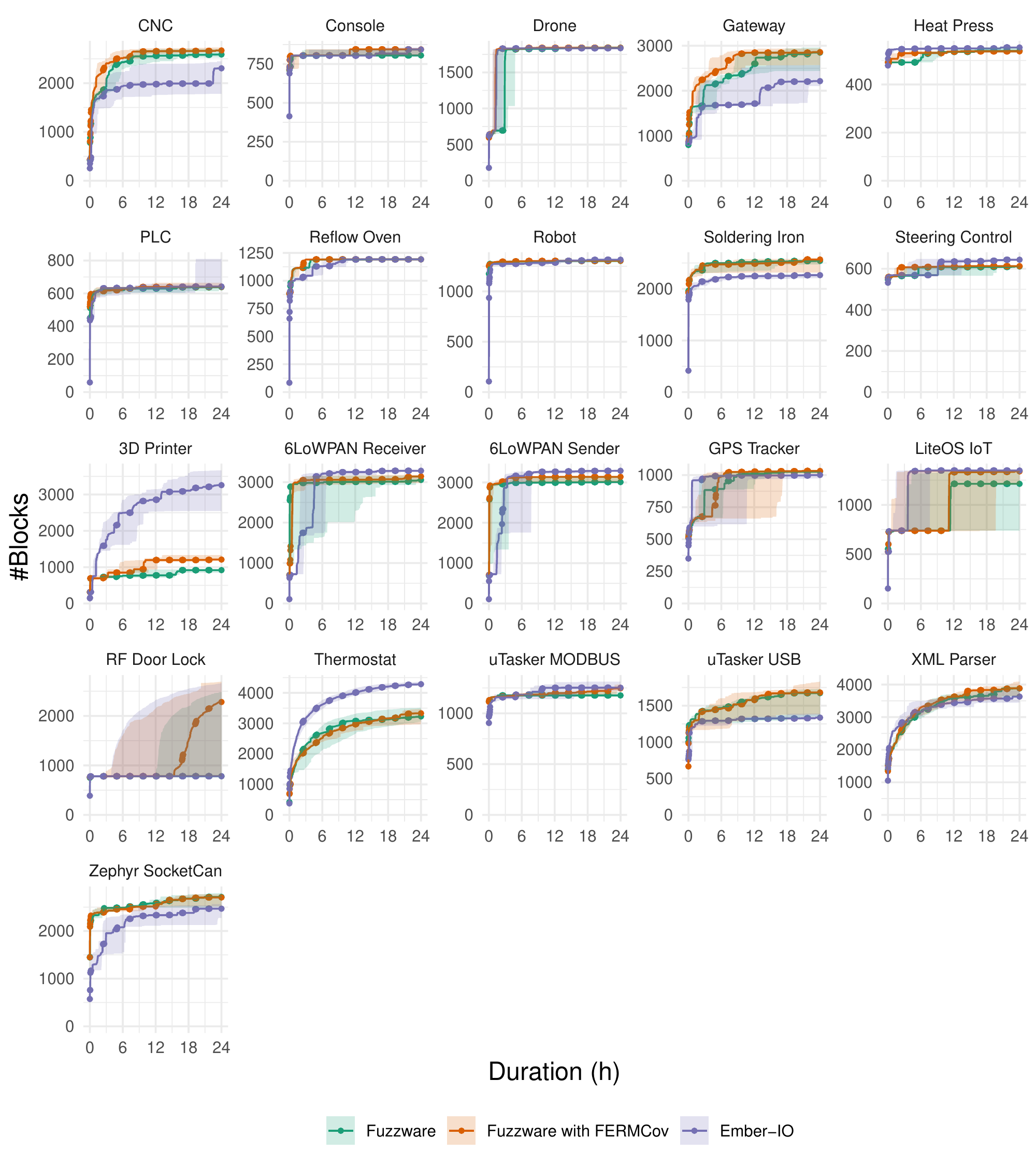}
    \caption{Comparison of coverage achieved over time by each fuzzer: {\sc Fuzzware}, \emu{} and {\sc Fuzzware} with our \fcov{} technique added. Bands represent the full range of coverage observed over the five trials performed for 24 hours on each firmware. \emu{} achieved higher coverage that was deemed statistically significant in the 3D Printer, 6LoWPAN Receiver, 6LoWPAN Sender, LiteOS IoT, Robot, Steering Control, Thermostat and uTasker MODBUS binaries (8 firmware). {\sc Fuzzware} achieved an improvement that was statistically significant in the CNC, GPS Tracker and Soldering Iron binaries (3 firmware). Applying \fcov{} to {\sc Fuzzware} yielded increases in coverage in the 3D Printer and uTasker MODBUS binaries that were statistically significant, with no significant reductions on any of the binaries.}
    \label{fig:cov_plot}
\end{figure*}
\clearpage
\subsection{Coverage Analysis}\label{apd:coverage-analysis}
Further investigating the blocks reached using \emu{} that were \textit{not} covered by {\sc Fuzzware}, we identified multiple sources for new coverage reached in the different binaries.
\begin{itemize}
\item In firmware such as the 3D Printer and 6LoWPAN Receiver binaries, coverage was increased through the exploration of more interesting data values and system states.
\item In firmware such as the Robot and LiteOS IoT binaries, we observed additional coverage in error handlers relating to timeouts. These timeouts depend on a long sequence of inputs without a checked bit being set. We consider peripheral input playback (PIP) an important component responsible for generating inputs that trigger timeouts.
\item Another interesting case is the Steering Control binary, where we observed many additional blocks being reached through the memory allocator. By using peripheral input playback to provide longer strings as serial commands, larger buffers must be allocated. This led to more code coverage within \textit{malloc}, \textit{realloc} and \textit{free}.
\end{itemize}

Further investigating the impact of \fcov{} when applied to {\sc Fuzzware}, in addition to the 3D Printer and uTasker MODBUS binaries where \fcov{} provided a statistically significant coverage improvement to {\sc Fuzzware}, we can observe an improvement in the speed of exploration. Numerous firmware depend on specific input sequences to complete initialisation, or reach key states required for proper firmware functionality, creating a hurdle for the fuzzer to overcome. After these difficult to solve sequences have been solved, essentially all of the blocks in the 24 hour run are reached and coverage plateaus; this is observable in Figure~\ref{fig:cov_plot}. While the addition of \fcov{} may have no significant impact on coverage after 24 hours in some binaries, it can reduce the time required to overcome these hurdles. For example: i)~in the Drone firmware, the median time to find an input that succeeds in initialisation and executes the main program loop was reduced from more than \textit{three hours} to approximately \textit{one and a half hours}; ii)~for the Heat Press binary, this time was reduced from more than \textit{six hours} to \textit{less than two hours}; iii)~for the Reflow Oven binary, \fcov{} reduced the time from \textit{four hours} to \textit{three hours}; and iv)~for the Steering Control binary, the time was reduced from \textit{five and a half hours} to \textit{less than two hours}. By using \fcov{} to remove functionally equivalent paths from the fuzzing queue, less time is spent mutating inputs that provide no new progress towards reaching the main firmware loop. This allowed us to reach the the main loop much faster in these firmware.

\subsection{Crash Analysis}\label{apd:crash-causes}
\begin{table}[h]
\centering
\caption{Classifications of crash root causes. The \textit{Replicated} row corresponds to known issues previously reported by the authors of {\sc Fuzzware}. The \textit{New} row corresponds to crashes not reported previously and using \emu{} techniques. Each root cause is categorised as a security issue, initialisation issue, or a false positive when the crash should not be trigger-able on the original device.}
\label{fig:crash_tab}
\resizebox{\columnwidth}{!}{
\begin{tabular}{|l|l|l|l|l|} 
\cline{2-5}
\multicolumn{1}{l|}{} & \multicolumn{1}{c|}{Security} & \multicolumn{1}{c|}{\begin{tabular}[c]{@{}c@{}}Uninitialised \\Variable Usage\end{tabular}} & \multicolumn{1}{c|}{\begin{tabular}[c]{@{}c@{}}False\\Positive\end{tabular}} & \multicolumn{1}{c|}{Total}  \\ 
\hline
\textit{Replicated} ({\sc Fuzzware})            & 17                            & 14                                                                                          & 1                                                                            & 32                          \\ 
\hline
\textit{New} (\emu{})                  & 2                             & 2                                                                                           & 2                                                                            & 6                           \\ 
\hline
Total                 & 19                            & 16                                                                                          & 3                                                                            & 38                          \\
\hline
\end{tabular}}
\end{table}
In addition to the 32 crashes we replicated from those reported in {\sc Fuzzware}~\cite{fuzzware}, 3 additional crashes were reported by the authors of {\sc Fuzzware} we discuss these below:
\begin{itemize}
    \item \textbf{uTasker MODBUS.~}During our investigations, we discovered that the 3 crashes in the uTasker MODBUS binary reported by {\sc Fuzzware} as three separate bugs result from the \textit{same uninitialised variable}---i.e. the same root cause. \emu{} correctly discovered one crash and consequently the root cause---the uninitialised variable. Although we have continued to report these three crashes as three separate bugs in Table~\ref{fig:crash_tab}, whether these are three different or the same software bug is an interesting question to consider. Additionally, {\sc Fuzzware} with \fcov{} identified another location within the uTasker MODBUS binary (not reported by the authors of {\sc Fuzzware}) where the use of the same uninitialised variable resulted in a new crash. However, we have chosen to not report this as a new bug in the total new bugs discovered by \emu{} based on our view that these are all the same bug resulting from the same root cause.
     
    \item \textbf{GPS Tracker.~}The third crash within the GPS Tracker binary is guarded by a string comparison. The {\sc Fuzzware} authors acknowledged the difficulty and non-deterministic nature of reproducing this crash, even with their fuzzing framework~\cite{fuzzware_gps_bug}. Because, AFL is not well suited to solving these types of comparisons.
    \item Interestingly, \textit{none} of these 3 additional crashes we discuss here were triggered in our 24 hour tests (involving 5 repeated runs) using {\sc{Fuzzware}}. However, the GPS Tracker crash, and one of the two missing uTasker MODBUS crashes, due to the uninitialised variable we discussed above, were observed during our tests using {\sc Fuzzware} with \fcov{}. This seems to confirm the effectiveness of the techniques we proposed as the bugs were reached more consistently across the fuzzing campaigns.
 
\end{itemize}
\vspace{1mm}
The two new security bugs (Gateway and Zephyr SocketCan binaries) identified using \emu{} (see Table~\ref{fig:crash_tab}) result from invalid bound checks on arrays. The triggering of these crashes is not dependent on new code coverage, instead requiring many iterations of the same code that loads data into the arrays. The use of \fcov{} assists in removing of test cases that are otherwise uninteresting, that the fuzzer would normally focus on due to the perceived new code coverage. This allows the fuzzer to, instead, prioritise increasing the number of iterations executed in existing code sections. Peripheral input playback simplifies the fuzzer mutations required in order to increase the amount of data stored in the buffers. We found our techniques particularly effective in helping to discover this \textit{class} of security bugs---i.e. buffer overflows.

We provide additional information regarding each bug, and seeds for reproducing these bugs within our repository at\linebreak \textit{\url{https://github.com/Ember-IO/Ember-IO-Experiments}}.

\end{document}